\documentclass[pre,aps,manuscript]{revtex4}

\usepackage{graphicx}
\usepackage{dcolumn}
\usepackage{bm}

\newcommand{\bs}{\bigskip}
\newcommand{\bc}{\begin{center}}
\newcommand{\ec}{\end{center}}

\begin{document}

\title{Non--Markovian diffusion over potential barrier 
in the presence of periodic time modulation}
\author{ V. M. \surname{Kolomietz} 
\thanks{Electronic address: vkolom@kinr.kiev.ua}}
\author{ S. V. \surname{Radionov} 
\thanks{Electronic address: Sergey.Radionov@matfys.lth.se}}
\affiliation{\textit{Institute for Nuclear Research, 03680 Kiev, Ukraine} }
\date{\today}

\begin{abstract}
The diffusive non--Markovian motion over a single--well potential barrier in
the presence of a weak sinusoidal time--modulation is studied. We found
non--monotonic dependence of the mean escape time from the barrier on a
frequency of the periodic modulation that is character to the stochastic
resonance phenomenon. The resonant acceleration of diffusion over the
barrier occurs at the frequency inversely proportional to the mean
first--passage time for the motion in the absence of the time--modulation.
\end{abstract}

\maketitle

%\keywords{ }

\section{Introduction}

The response of complex nonlinear systems on periodic time input may have
the features that are absent for linear systems. The very famous example of
such features is the stochastic resonance phenomenon \cite{sr1}, when the
response of the nonlinear system on the harmonic perturbation is resonantly
activated under some optimal level of a noise. The resonant activation of
the system occurs when the frequency of the modulation is near the Kramers'
escape rate of the transitions from one potential well to another. The
stochastic resonance phenomenon has been found and studied in many physical
systems like a ring laser \cite{sr11}, magnetic systems \cite{sr12}, optical
bistable systems \cite{sr13} and others. The prototype of the stochastic
resonance studies is a model of overdamped motion between potential wells
of the bistable system. The frequency of the transitions between wells is
given by the famous Kramers' rate and the stochastic resonance is achived
when a frequency of an external periodic modulation is of the order of the
Kramers' rate.

In the present paper, we would like to study a diffusion over a single--well
potential barrier in the presence of periodic time modulation. The diffusion
over the barrier is generated by a colored noise whose statistical properties
are related to the retarded friction term. In addition to the works
\cite{colnois1}--\cite{colnois3}, we are going to study how the correlation
time of the colored noise influence the first passage time distribution and 
escape rate over the barrier. 

The paper is organized as follows. In Sect.~\ref{dmopb}, we set in basic
Langevin equation of motion for diffusive motion over a potential barrier in
the presence of sinusoidal time modulation. In Sect.~\ref{sfb}, it is
considered the unperturbed path of the model system. The
time modulated diffusion is discussed in Sect.~\ref{pmfb}. 
Finally, the main conclusions of the paper are given in Sect.~\ref{concl}.

\section{Diffusive motion over a potential barrier}

\label{dmopb}

We start from a quite general Langevin formulation of the problem of
diffusive overcoming of a potential barrier in the presence of a harmonic
perturbation $V_{\mathrm{ext}}(t)=\alpha \cdot \mathrm{sin}(\omega t)$:

\begin{equation}
M\ddot{q}(t)=-\frac{\partial E_{\mathrm{pot}}}{\partial q}
-\int_{0}^{t}\kappa (t-t^{\prime })\dot{q}(t^{\prime })dt^{\prime }+\xi
(t)+\alpha \cdot \mathrm{sin}(\omega t),  \label{lang1}
\end{equation}%
where $q$ is the dimensionless coordinate, $M$ is the mass, $\kappa
(t-t^{\prime })$ is the memory kernel and $\xi (t)$ is the random force. The
potential energy $E_{\mathrm{pot}}$ is schematically shown in \textrm{Fig.~1
}and presents a single--well barrier formed by a smoothing joining at 
$q=q^{\ast }$ of the potential minimum oscillator with the inverted
oscillator,
\begin{eqnarray}
E_{\mathrm{pot}}=\frac{1}{2}M\omega _{A}^{2}(q-q_{A})^{2},~~~~q\leq q^{\ast },  
\nonumber \\
=E_{\rm pot,B}-\frac{1}{2}M\omega _{B}^{2}(q-q_{B})^{2},~~~~q>q^{\ast }.  
\label{Epot}
\end{eqnarray}

A noise term $\xi (t)$ in Eq.~(\ref{lang1}) is assumed to be Gaussian
distributed with zero mean and correlation function related to a memory
kernel $\kappa (t-t^{\prime })$ of a retarded friction force:

\begin{equation}
\langle \xi (t)\ \xi (t^{\prime })\rangle =T\ \kappa (t-t^{\prime }).
\label{flucdis}
\end{equation}
Below we shall assume that the memory kernel is given by

\begin{equation}
\kappa (t-t^{\prime })=\kappa _{0}\mathrm{exp}\left( -\frac{|t-t^{\prime }|}{%
\tau }\right) ,  \label{kappa}
\end{equation}
where $\tau $ is a correlation time.

\section{Numerical calculations}

\label{numeric}

In the numerical calculations, we have measured all quantities 
quantities of the dimension of energy in units of the temperature of the
system $E_0=T$, quantities of the dimension of time in units of 
$t_0=\sqrt{M/T}$ and quantities of the dimension of frequency in
units of $\omega_0=\sqrt{T/M}$. For the system's parameters   
$q_A,~q^*,~q_B,\omega _{A},~\omega _{B},~E_{\rm pot,B}$ and$~\kappa _{0}$,
we adopted the values:
\begin{eqnarray}
q_{A}=1,~~q^{\ast}=1.2,~~q_{B}=1.6,
\nonumber\\
\omega_A=6.75,~~\omega_B=9.59,~~
E_{\rm pot,B}=5.15,~~\kappa_{0}=1920.  
\label{parameters}
\end{eqnarray}
which are widely used under model diffusion--like studies of fission of
highly excited atomic nuclei, see Ref.~\cite{kora01}.

\subsection{Unperturbed diffusion over the barrier}

\label{sfb}

At the beginning, we investigated the non--Markovian diffusive dynamics for
the infinitely slow ($\omega =0$) modulation and calculated a first--passage
time distribution. For that, the Langevin equation (\ref{lang1}) was solved
numerically by generating a bunch of the trajectories, all starting at the
potential well (point A in \textrm{Fig.~1}) and having the initial
velocities distributed according to the Maxwell--Boltzman distribution.
First, we would like to study the diffusive dynamics 
(\ref{lang1})--(\ref{kappa}) in terms of first passage time distribution.

In \textrm{Fig.~2}, the mean first--passage time is presented as a function
of the correlation time $\tau $. 

An increase of the mean first--passage time $t_{\mathrm{mfpt}}$ with the
relaxation time $\tau $ means that memory effects in the Langevin dynamics 
(\ref{lang1}) hinders the diffusion over the barrier. A non--monotonic growth
of the mean first--passage time is caused by transition (occurred at the 
$\tau \approx 0.007~$) from nearly Markovian regime of the
diffusion (\ref{lang1}) to the regime where the memory effects are quite
important. As far as the memory effects become stronger and stronger ($\tau
\rightarrow \infty $), the value of $t_{\mathrm{mfpt}}$ reaches a finite
limit. Here, the hindrance of the escape over the barrier is caused by the
renormalization of the ordinary conservative force in Eq.~(\ref{lang1} )
that obtains an additional contribution from the time--retarded force

\begin{equation}
-\kappa _{0}\int_{0}^{t}\mathrm{exp}\left( -\frac{|t-t^{\prime }|}{\tau }%
\right) \dot{q}(t^{\prime })dt^{\prime }\rightarrow -\kappa _{0}\left[
q(t)-q_{A}\right] ,~~~~~\tau \rightarrow \infty .  \label{tauinf}
\end{equation}

In the opposite limit of quite small values of the correlation time $\tau$, the
hindrance is exclusively due to an usual friction:

\begin{equation}
-\kappa _{0}\int_{0}^{t}\mathrm{exp}\left( -\frac{|t-t^{\prime }|}{\tau }%
\right) \dot{q}(t^{\prime })dt^{\prime }\rightarrow -\kappa _{0}\tau \dot{q}%
(t),~~~~~\tau \rightarrow 0.  \label{tau0}
\end{equation}%
In intermediate case, the hindrance is defined both by the friction and the
additional elastic force,

\begin{equation}
-\kappa _{0}\int_{0}^{t}\mathrm{exp}\left( -\frac{|t-t^{\prime }|}{\tau }%
\right) \dot{q}(t^{\prime })\ dt^{\prime }=-\gamma (t,\tau )\ \dot{q}%
(t)-C(t,\tau )\ q(t),  \label{split}
\end{equation}%
where the effective friction coefficient $\gamma (t,\tau )$ may be quite
well approximated by

\begin{equation}
\gamma (t,\tau )\approx \frac{\kappa _{0}\cdot \tau }{1+(\kappa _{0}/M)\tau
^{2}},~~~~~t\gg \tau .  \label{gamma}
\end{equation}%
The stiffness parameter $C(t,\tau )$ in Eq. (\ref{split}) grows from $0$ to $%
\kappa _{0}$ with a growing of $\tau $.

Secondly, we measure the diffusion dynamics (\ref{lang1})--(\ref{kappa})
through an escape rate, $R(t)$, characteristics. The escape rate over the 
barrier is defined as follows

\begin{equation}
R(t)=-\frac{1}{P(t)}\frac{dP(t)}{dt},  \label{r}
\end{equation}%
where $P(t)$ is the survival probability, i.e., probability of 
finding the system on the left from the top of the barrier up to time $t$:

\begin{equation}
P(t)=N(t)/N_{0}.  \label{Pt}
\end{equation}

Here, $N(t)$ is a number of the trajectories being not reaching the top of
the barrier up to time $t$ and $N_{0}$ is a total number of the trajectories
involved in the calculations. In \textrm{Fig.~3}, it is plotted the typical
time behavior of the escape rate $R(t)$ for quite small $\tau =0.005$ and 
fairly large $\tau =0.026$ values of the correlation time.

It is seen from \textrm{Fig. 3} that initially the escape is affected by
transient effects, when the survival probability $P(t)$ deviates strongly
from the exponential form. With time, the escape process becomes more and
more stationary giving rise to the corresponding saturation of the rate 
$R(t) $ (\ref{r}) establishing of a quasistationary probability flow over the
barrier. Qualitatively, one can describe typical time evolution of the
escape rate as

\begin{equation}
R(t)=R_{0}(1-\mathrm{e}^{-t/t_{\mathrm{tran}}}).  \label{r0}
\end{equation}%
In both cases a duration of the transient period, $t_{\mathrm{tran}}$, is
almost the same ($t_{\mathrm{tran}}\approx 50$) for
quite weak and fairly large memory effects in the diffusion process.
However, a saturation value, $R_{0}$, of the escape rate is significantly
different. It is because of the memory effect for the large values of the
correlation time $\tau $.

In \textrm{Fig.~4}, we showed how the value $R_{0}$ (\ref{r0}) depends on
the size of the memory effects in the diffusive dynamics (\ref{lang1}).
Dotted line in \textrm{Fig. 4} represents the famous Kramers' result for the
escape rate value in a quasistationary regime \cite{Kramers}

\begin{equation}
R_{Kr}=\frac{\omega _{A}}{2\pi }\left( \sqrt{1+\left[ \frac{\gamma (\tau )}{%
2M\omega _{B}}\right] ^{2}}-\frac{\gamma (\tau )}{2M\omega _{B}}\right)
\mathrm{exp}\left( -\frac{B}{T}\right) ,  \label{Kramers}
\end{equation}
where the $\tau $--dependent friction coefficient $\gamma (\tau )$ is given
by Eq.~(\ref{gamma}).

We see that the memory effects significantly suppress the value of the
escape rate in the saturation regime of probability flow over the potential
barrier. Initially (i. e., at relatively small values of the correlation time 
$\tau $) the suppression is mainly caused by the growing role of the usual
friction in the non--Markovian diffusion (\ref{lang1})-- (\ref{kappa}), see
also Eqs.~(\ref{split}) and (\ref{gamma}). As is followed from the \textrm{%
Fig. 4}, in this case the escape rate at saturation $R_{0}$ (\ref{r0}) may
be quite well approximated by the Kramers' formula (\ref{Kramers}). On the
other hand, at relatively large correlation times $\tau $, the effect of the
friction on the diffusion over the barrier is negligibly weak and the escape
rate's suppression appears exclusively due to the additional conservative
force, see Eq. (\ref{split}). As a result, the stationary value of the
escape rate deviates substantially from the Kramers' escape rate (\ref%
{Kramers}) at the fairly strong memory effects in the diffusive motion
across the barrier.

\subsection{Diffusion over the barrier in the presence of a periodic
modulation}

\label{pmfb}

Now we will study the diffusion over the barrier (\ref{lang1})--(\ref{kappa})
in the presence of the external harmonic force. We will assume that the
amplitude $\alpha $ of the force $\alpha \cdot \mathrm{sin}(\omega t)$ in
Eq. (\ref{lang1}) is so small ($\alpha =0.05$) that still the
reaching the top of the barrier is caused exclusively by diffusive nature of
the process.

In \textrm{Fig.~5}, we calculated the typical dependencies of the mean
first--passage time $t_{\mathrm{mfpt}}$ on the frequency $\omega $ of the
external harmonic force. The calculations were performed for the weak, $\tau
=0.005$, (lower curve in \textrm{Fig. 5}) and strong, 
$\tau =0.026$, (upper curve in \textrm{Fig. 5}) memory
effect on the non--Markovian diffusive motion over the barrier. 

In both cases the mean first--passage time $t_{\mathrm{mfpt}}$
non--monotonically depends on the frequency of the perturbation that is
character for the stochastic resonance phenomenon observed in a number of
different physical systems. From \textrm{Fig. 5} one can conclude that
diffusion over the potential barrier in the presence of the harmonic time
perturbation is maximally accelerated at some definite so to say resonant
frequency $\omega _{\mathrm{res}}$ of the perturbation,

\begin{equation}
\omega _{\mathrm{res}}\approx \frac{1.5}{t_{\mathrm{mfpt}}(\omega =0)}
\label{resonance}
\end{equation}%
see also \textrm{Figs. 2} and \textrm{4}. In fact, the quantity $t_{\mathrm{%
mfpt}}(\omega =0)$ presents the characteristic time scale for the diffusion
dynamics (\ref{lang1}). In the case of adiabatically slow time variations of
the harmonic force, $\omega t_{\mathrm{mfpt}}(\omega =0)\ll 1$ and $t<t_{%
\mathrm{mfpt}}$, one can approximately use $\alpha \cdot \mathrm{sin}(\omega
t)\approx \alpha \cdot \omega t$ and the diffusion over the barrier is
slightly accelerated. As a result of that, the mean first--passage time $t_{%
\mathrm{mfpt}}(\omega )$ is smaller than the corresponding unperturbed value
$t_{\mathrm{mfpt}}(\omega =0)$. The same feature is also observed at the
fairly large modulation's frequencies. Thus, in the case of $\omega t_{%
\mathrm{mfpt}}(\omega =0)\gg 1$, the harmonic perturbation $\alpha \cdot
\mathrm{sin}(\omega t)$ may be treated as a random noise term with the zero
mean value and variance $\alpha ^{2}$. Such a new stochastic term will lead
to additional acceleration of the diffusion over the barrier.

The existence of the resonant regime (\ref{resonance}) in the periodically
modulated diffusion process (\ref{lang1}) is even more clear visible in the
escape rate characteristics of the process. We have plotted in \textrm{Fig.~6%
} the time evolution of the escape rate (\ref{r}) found for the resonant
frequency $\omega _{\mathrm{res}}$ (\ref{resonance}) (curve 1 in \textrm{%
Fig. 6}), the quite smaller frequency $\omega =\omega _{\mathrm{res}}/10$
(curve 2 in the \textrm{Fig. 6}) and the quite larger frequency $\omega =10$
$\omega _{\mathrm{res}}$ (curve 3 in \textrm{Fig. 6}) of the modulation.

Again, at very slow ($\omega =\omega _{\mathrm{res}}/10$) and fast ($\omega
=10\omega _{\mathrm{res}}$) perturbations the escape rate $R(t)$ looks very
similar to the corresponding unperturbed value $R(t,\omega =0)$, compare
\textrm{Figs.~3} and \textrm{6}. In other words, the initial transient
period in the time evolution of the escape rate is followed by the
stationary regime, when the escape subsequently saturates with time.
Contrary to that, the escape rate shows complex time behavior as far as the
periodic modulation occurs at the resonant frequency ($\omega =\omega _{%
\mathrm{res}}$), see curve 1 in \textrm{Fig.~6}. This resonant regime of the
modulated diffusion is essentially nonstationary when the system remains
excited during quite long time. We checked such a feature for larger number
of trajectories and longer time intervals used to calculate the
escape rate characteristics (\ref{r}).

\section{Conclusions}

\label{concl}

In the present study, we have investigated how model non--Markovian 
diffusion over the single--well parabolic barrier is affected by the
external periodic time modulation. We have calculated both the mean 
first--passage time $t_{\mathrm{mfpt}}$ and
the escape rate $R(t)$ over the barrier. These two quantities have been
found to be sensitive to the relative size of memory effects in the
diffusive dynamics (\ref{lang1})--(\ref{kappa}), measured by the correlation time
$\tau $. Thus, we have demonstrated that the memory effects hinder the
escape over the barrier, see \textrm{Figs.~2 }and\textrm{\ 4}. In contrast
to the motion in the presence of usual friction force, the hindrance of the
escape occurs due to the Markovian friction and additional conservative
components (\ref{split}) of the retarded time force in Eq.~(\ref{lang1}).
Having calculated the mean first--passage time $t_{\mathrm{mfpt}}$ for
different values of the frequency $\omega $ of the modulation, we have found
that the sinusoidal perturbation accelerates the diffusion over the barrier,
see \textrm{Fig.~5}. The maximal (resonant) acceleration is achieved at the $%
\omega =\omega _{\mathrm{res}}$, where $\omega _{\mathrm{res}}$ is inversely
proportional to the mean first--passage time in the absence of the
modulation (\ref{resonance}). We have seen that a value of the resonant
activation over the barrier $t_{\mathrm{mfpt}}(\omega =\omega _{\mathrm{res}%
})/t_{\mathrm{mfpt}}(\omega =0)$ remains practically the same for the quite
weak as well as for the fairly strong memory effects in the diffusive
dynamics. It has been observed that the diffusive dynamics (\ref{lang1})--(%
\ref{kappa}) in the resonant activation regime (\ref{resonance}) has the
peculiarity, reflecting in the complex time behavior of the escape rate $%
R(t) $, see \textrm{Fig.~6}. Importantly, that the absence of the escape
rate's saturation with time implies essentially transient character of the
events of the first passing the top of the potential barrier.

\newpage

\bc

Figure captions.

\ec

\bs
\bs

Fig.~1. Landscape of potential energy $E_{\mathrm{pot}}(q)$ of
Eq.~(\protect\ref{Epot}).

\bs

Fig.~2. The mean first--passage time $t_{\mathrm{mfpt}}$ of the
non--Markovian diffusion process (\protect\ref{lang1})--(\protect\ref{kappa})
vs the correlation time $\protect\tau$.

\bs

Fig.~3. The time dependence of the
escape rate $R(t)$ (\protect\ref{r}) of the non--Markovian diffusion process
(\protect\ref{lang1})-- (\protect\ref{kappa}) calculated for the cases of
quite small $\protect\tau =0.005$ and fairly large
$\protect\tau =0.026$ values of the correlation time.

\bs

Fig.~4. The saturation value $R_{0}$ of the escape rate
(\protect\ref{r})--(\protect\ref{r0}) vs the
strength $\protect\tau $ of the memory effects in the non--Markovian
diffusion process (\protect\ref{lang1})--(\protect\ref{kappa}). The dotted
line represents the Kramers' result (\protect\ref{Kramers}) for the escape
rate calculated with the $\protect\tau $--dependent friction coefficient of
Eq. (\protect\ref{gamma}).

\bs

Fig.~5. The mean first--passage time $t _{\mathrm{mfpt}}$ of the non--Markovian
diffusion process (\protect\ref{lang1})--(\protect\ref{kappa}) is given as
a function of the frequency $\protect\omega $ of the harmonic time perturbation
at two values of the correlation time $\protect\tau =0.005$
(lower curve) and $\protect\tau =0.026$ (upper curve).

\bs

Fig.~6. The time dependence of the escape
rate $R$ (\protect\ref{r}) for the periodically modulated diffusion over the
barrier (\protect\ref{lang1})--(\protect\ref{kappa}) are shown at the
resonant frequency $\protect\omega _{\mathrm{res}}$
(\protect\ref{resonance}) (curve 1), fairly smaller frequency
$\protect\omega =\protect\omega _{\mathrm{res}}/10$ (curve 2) and quite
bigger frequency $\protect\omega =10\protect\omega _{\mathrm{res}}$ (curve 3)
of the periodic modulation. The dependencies were calculated for the memory
time $\protect\tau =0.01$.

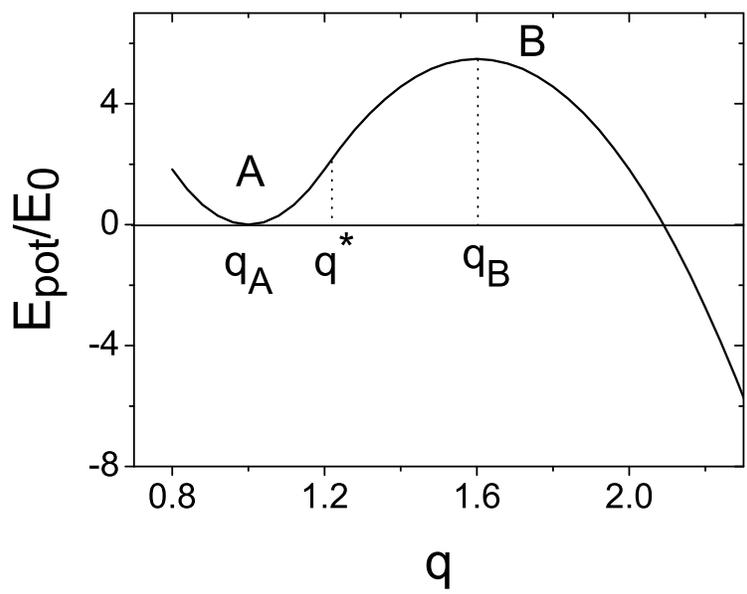
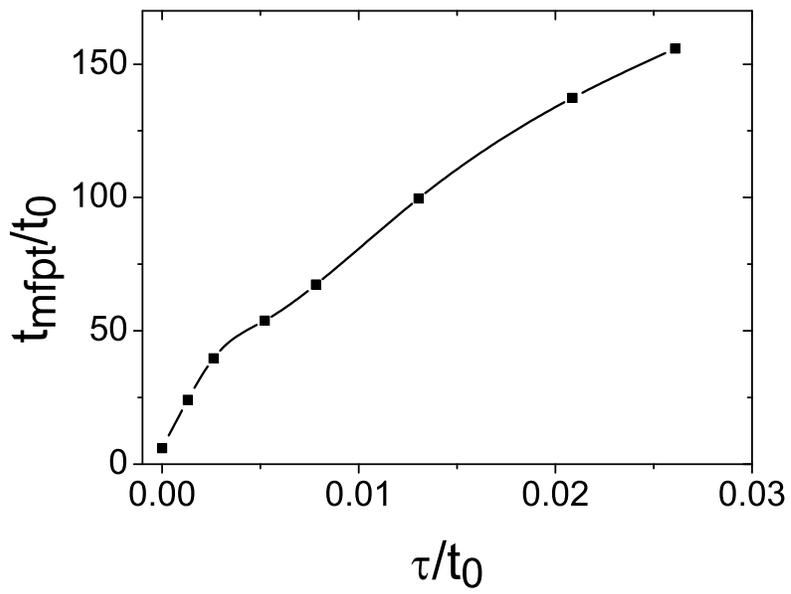
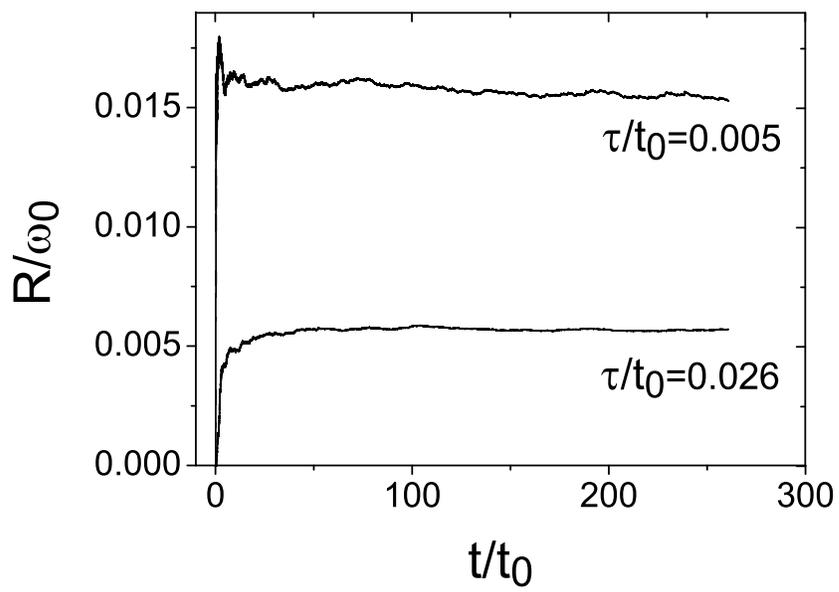
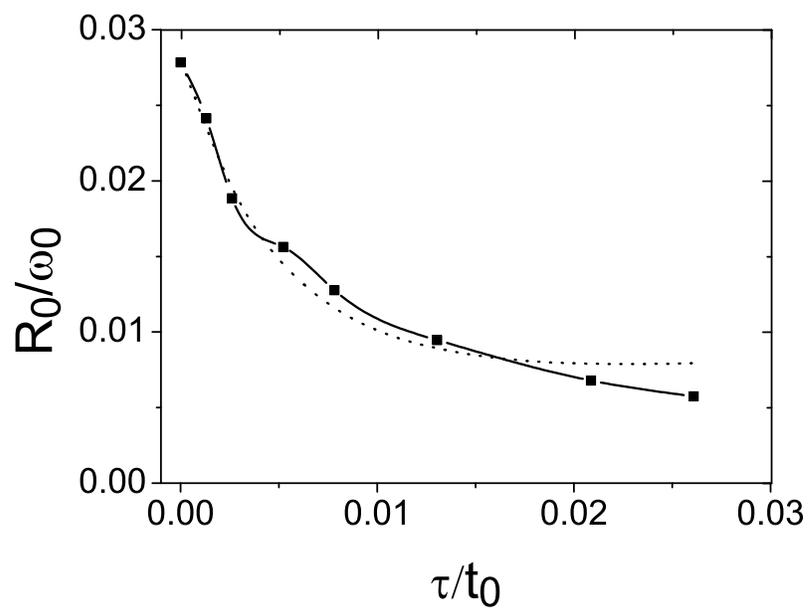
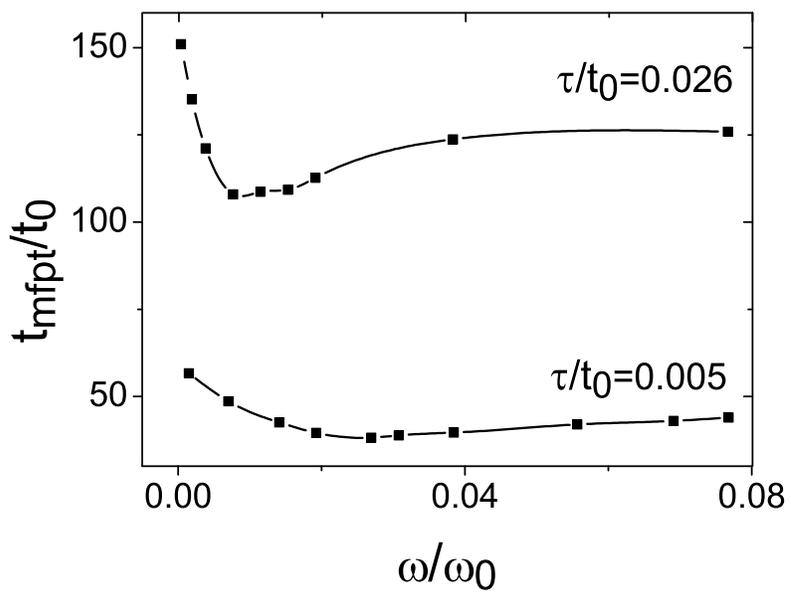
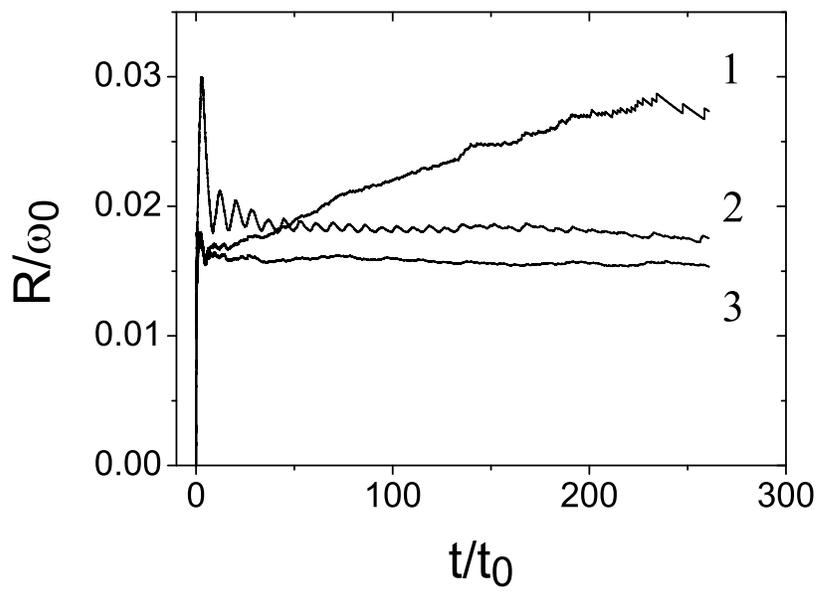
\end{document}